\documentclass[%
 reprint,
superscriptaddress,
 amsmath,amssymb,
 prl,
]{revtex4-1}

\usepackage{graphicx}
\graphicspath{{./fig/}}
\usepackage{dcolumn}
\usepackage{bm}

\begin{document}

\preprint{APS/123-QED}

\title{Scale invariance at the onset of turbulence in Couette flow}

\author{Liang Shi}
\email{Email address: liang.shi@ds.mpg.de}
\affiliation{%
  Max Planck Institute for Dynamics and Self-Organization (MPIDS), 37077 G\"ottingen, Germany
}
\affiliation{Institute of Geophysics, University of G\"ottingen, 37077 G\"ottingen, Germany}

\author{Marc Avila}
\email{Email address: mavila@lstm.uni-erlangen.de}
\affiliation{%
 Max Planck Institute for Dynamics and Self-Organization (MPIDS), 
 37077 G\"ottingen, Germany
}
\affiliation{%
 Institute of Fluid Mechanics, Friedrich-Alexander-Universit\"at Erlangen-N\"urnberg, 
 91058 Erlangen, Germany
}

\author{Bj\"orn Hof}
\email{Email address: bjoern.hof@ds.mpg.de}
\affiliation{%
  Max Planck Institute for Dynamics and Self-Organization (MPIDS), 37077 G\"ottingen, Germany
}
\affiliation{%
  Institute of Science and Technology Austria, 3400 Klosterneuburg, Austria
}

\date{\today}

\begin{abstract}
Laminar-turbulent intermittency is intrinsic to the transitional regime of a wide range of fluid flows including pipe, channel, boundary layer and Couette flow. In the latter turbulent spots can grow and form continuous stripes, yet in the stripe-normal direction they remain interspersed by laminar fluid. We carry out direct numerical simulations in a long narrow domain and observe that individual turbulent stripes are transient. In agreement with recent observations in pipe flow we find that turbulence becomes sustained at a distinct critical point once the spatial proliferation outweighs the inherent decaying process. By resolving the asymptotic size distributions close to criticality we can for the first time demonstrate scale invariance at the onset of turbulence.

\end{abstract}

\maketitle

Turbulence often arises despite the linear stability of the laminar flow and the nature of this transition has remained unresolved for over a century~\cite{Reynolds_PRSL1883}. Prominent examples are pipe, channel and Couette flows and here finite amplitude perturbations can lead to sufficiently strong distortions of the base flow, that for large enough Reynolds numbers ($Re=UL/\nu$, characteristic length $L$, velocity $U$ and kinematic viscosity $\nu$) causes transition to turbulence. At moderate $Re$ turbulent structures are localized and are commonly referred to as spots or stripes in Channel and Couette flow and puffs in pipe flow. Individual localized structures are of transient nature~\cite{Hof_NATURE2006,*Hof_PRL2008,*Echeverry_PRE2010,Avila_JFM2010} but they can also temporarily grow and seed new spots in their vicinity, overall resulting in complex spatio-temporal dynamics. As shown for pipe flow, this spreading rate increases with $Re$, eventually outweighing the decay~\cite{Avila_Science2011}. It has been argued that this mechanism causes a phase transition in the thermodynamic limit from transient to sustained turbulence~\cite{Moxey_PNAS2010}. As $Re$ surpasses the critical point, turbulence can overall survive due to the spatial spreading. The observed dynamics bears resemblance to contact process such as directed percolation (DP)~\cite{Pomeau_PhyD1986}. In this analogy laminar flow corresponds to the absorbing passive state and turbulence to the active one. However, for pipe flow~\cite{Avila_Science2011} the phase transition has not been characterized directly but has only been inferred from the statistical behaviour (mean splitting and decay times) of individual spots. The main difficulty in pipe flow is that the relevant time scales of the relevant processes are extremely large, putting a study of correlation exponents presently beyond reach. Here we report a numerical (direct numerical simulation, DNS) study of another fundamental shear flow, plane Couette flow (pCf) where the fluid is sheared between two sliding plates (see Fig.~\ref{geometry}(a)). As will be shown below, the timescales close to the critical point are much smaller than those in pipe flow. Hence size distributions and turbulent fractions can be determined close to criticality, which provides an unique opportunity to characterize the phase transition.

Because of its unlimited width in the spanwise direction ($z'$ in Fig.~\ref{geometry}), pCf is spatially more complex than pipe flow. In order to simplify the spatial-temporal complexity and to enable us to resolve the long interaction times, we choose a slender computational domain inclined with the flow streamwise direction, as shown in Fig.~\ref{geometry}(b). The idea of such a ``tilted'' computational domain was originally introduced to reduce the computing cost by Barkley and Tuckerman~\cite{Barkley_PRL2005,*Barkley_pof2011}. 

\begin{figure}[!ht]
  \begin{center}
    \includegraphics[width=.35\textwidth]{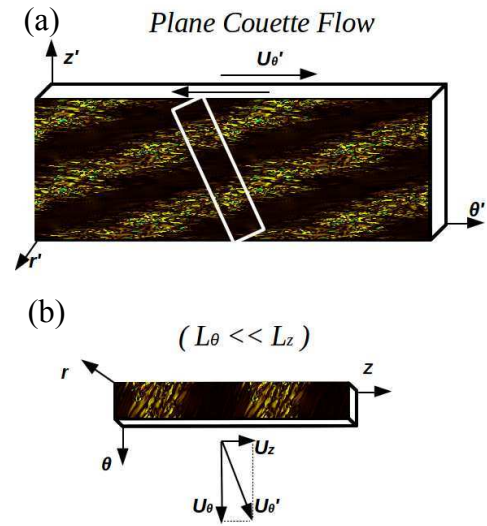}
  \end{center}
  \caption{Schematic of the computational domain for pCf. (a) normal box; (b) ``tilted'' slender box (white box in (a)). The stripe pattern in the background is a visualization of the streamwise vorticity field.}
  \label{geometry}
\end{figure}

Simulations are carried out using a numerical code for Taylor-Couette flow (flow between two rotating concentric cylinders)~\cite{Meseguer_EPJ2007}. The two cylinders are chosen to counter-rotate and a radius ratio (inner to outer cylinder radius) of $\eta=0.993$ has been selected so that curvature effects are negligible. As shown in an earlier study~\cite{Faisst_PRE2000} Taylor-Couette flow approaches the pCf limit for these parameter settings and is linearly stable for all $Re$ relevant to this study. By using the half gap distance $h$ and $\tau_h=h/U$ (here $U$ is the magnitude of the boundary velocity.) as the length and time unit, the dimensionless imcompressible Navier-Stokes equations are numerically solved with a Petrov-Galerkin pseudo-spectral method~\cite{Meseguer_EPJ2007}. We discretize the equations using Fourier modes in the in-plane directions and by modified Chebyshev polynomials in the wall-normal direction. The Reynolds number is here defined as $Re=Uh/\nu$ and the velocity field is decomposed into base flow and perturbation, $\mathbf{u} = \mathbf{U_{b}}+\mathbf{u'}$, where the perturbation part $\mathbf{u'}$ can be expressed as 
\begin{equation}
  \textbf{u}'(r,\theta,z)=\sum_{l=-L}^{L}\sum_{n=-N}^{N}\sum_{m=0}^{M}a_{lnm}e^{i(lk_z z+nk_{\theta} \theta)}\textbf{v}_{lnm}(r),
\label{vel}
\end{equation}
where $a_{lnm}$ are the spectral coefficients and $\textbf{v}_{lnm}$ are the Chebyshev polynomials. Periodic boundary conditions are imposed in the in-plane directions, while no-slip boundary conditions are applied in the wall-normal direction. The size of the computational domain is chosen sufficiently big in the $z$ direction $(L_z\times L_{\theta} \times L_r)=(100h\times 10h\times 2h)$ to study the evolution of isolated stripes whereas for the study of stripe interactions $L_z$ is increased to $960h$. The spatial resolution (number of grid points) for the smaller domain is $(2L\times 2N\times (M+1))=(512\times 48\times 27)$, the adequacy of which is checked by the one-dimensional energy spectra and by the convergence study of the statistical lifetime and splitting time distributions. For the larger domain the resolution in $z$ direction is increased to $6144$.

For the investigation of the decay and splitting of individual localized turbulent stripes, uncorrelated velocity fields of single stripes were used as initial conditions. These were generated by first simulating a fully turbulent flow at $Re=400$, followed by a sudden reduction to the desired Reynolds number $Re=325$. Since the decay and splitting are observed to depend strongly on the initial conditions, many realisations are necessary to determine the mean lifetimes and splitting times. The Reynolds numbers of interest are in the range of $Re\in [310,350]$. At each Re, $300$ realisations with different initial conditions are conducted and each simulation was run for a predefined time duration (cut-off time).
If a decay or splitting occurred earlier, the run was terminated after this event. Let $N^s$, $N^d$ and $N^c$ denote the number of splitting, decaying cases and the ones reaching the cut-off time, respectively.
By sorting in increasing order all the final times accociated with spliting events (or splitting time), we obtain a splitting time series $\{t_i^s\}_{i=1}^{N^s}$, with the probability that a stripe has not split up to a time $t$
\begin{equation}
  P(\text{splitting at }t\geqslant t_i^s)=P_i^s=1-(i-1)/N^s, \: i=1,\cdots,N^s.
\end{equation} 

\begin{figure}[!ht]
  \begin{center}
    \includegraphics[width=.45\textwidth]{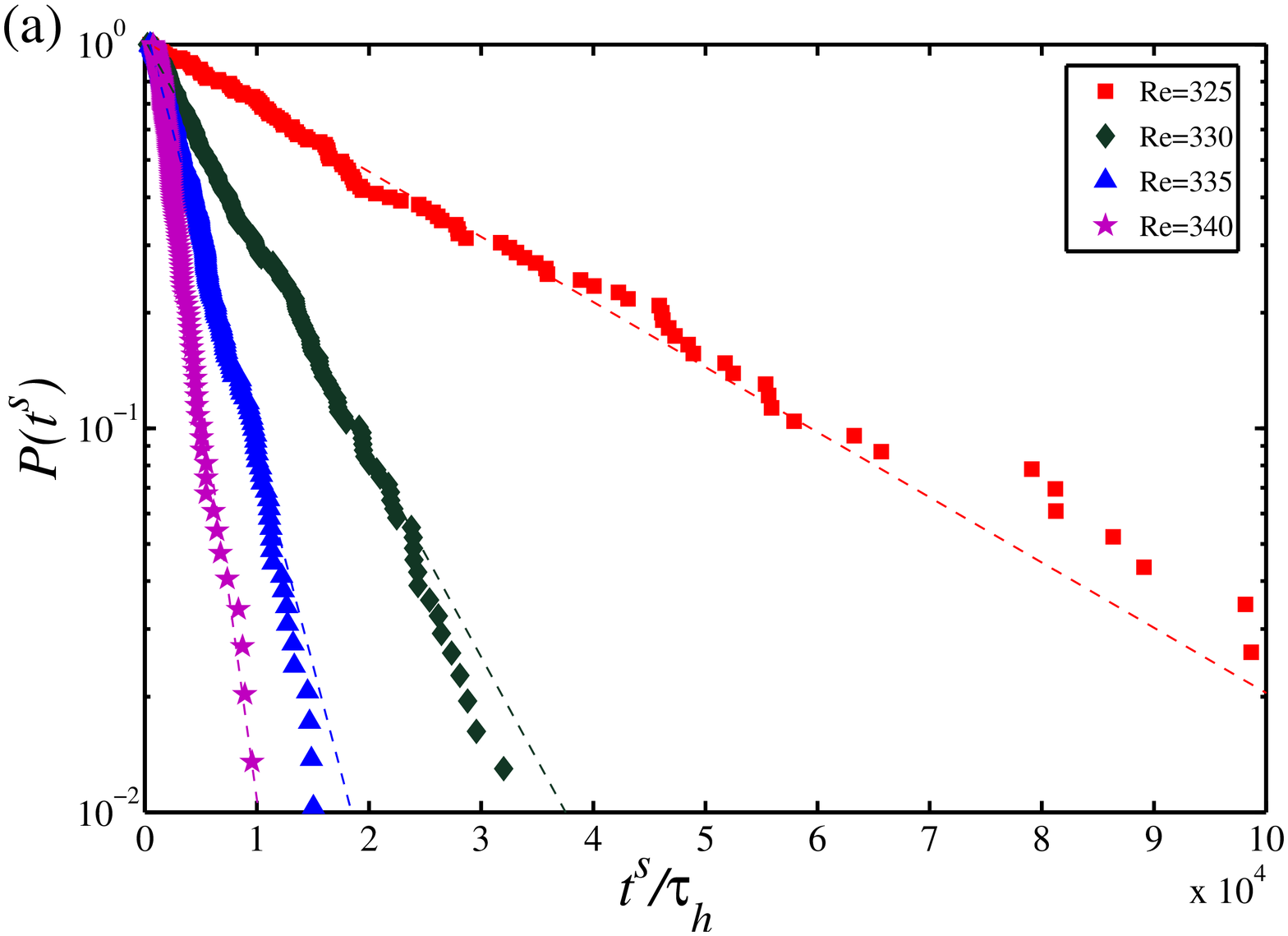}\\
    \includegraphics[width=.45\textwidth]{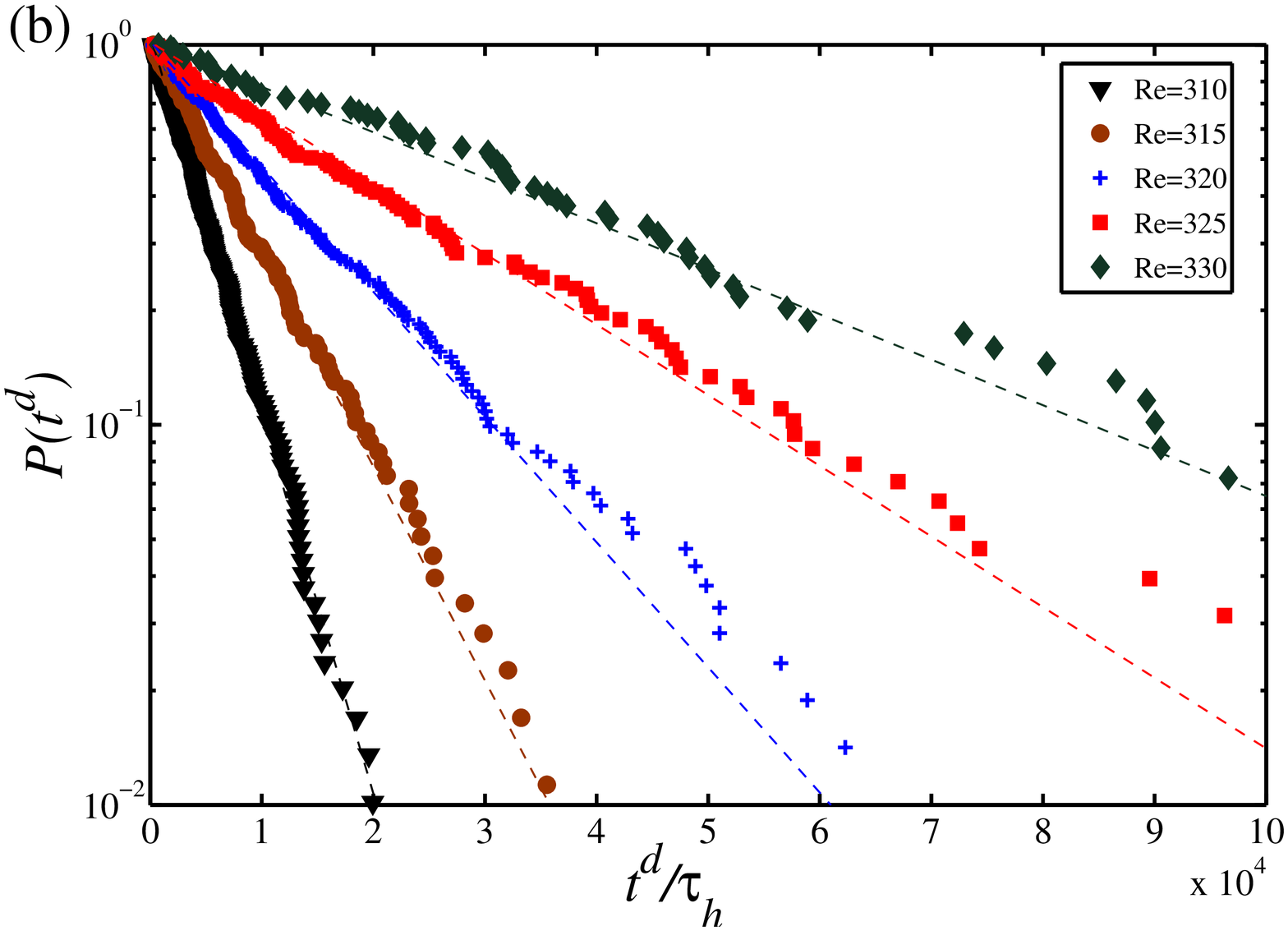}
  \end{center}
  \caption{Probability distributions of single turbulent stripe at different Reynolds numbers. (a) Splitting times; (b) lifetimes, in the unit of $\tau_h=h/U$. The dashed lines are the corresponding exponential curves from equation $(\ref{eq:expDist})$. Y-axis is in the logarithmic scale.}
  \label{fig:probdist}
\end{figure}

\begin{figure}[!ht]
  \begin{center}
    \includegraphics[width=.45\textwidth]{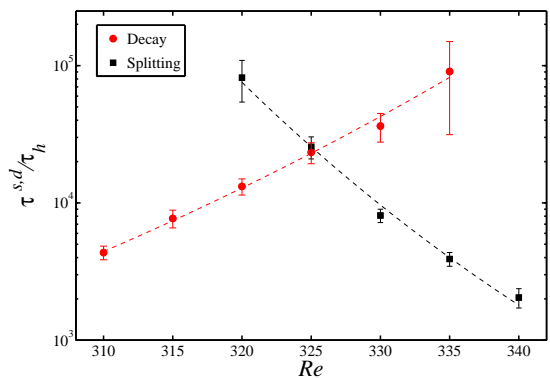}
  \end{center}
  \caption{Scaling of the parameter $\tau^{s,d}$ with $Re$ of the splitting (dark square) and decaying (red circle) of individual turbulent stripes. The error bars correspond to the 95\% confidential intervals. The dashed lines are the superexponential curves to guide the eyes.}
  \label{fig:tau}
\end{figure}

As shown in Fig.~\ref{fig:probdist}(a), the probability distributions of the splitting have exponential tails (excluding the initial formation period $t_0$). This shows that stripe splitting is a memoryless process and we can therefore determine the mean time $\tau^s(Re)$ for a splitting to occur by the following exponential ansatz, 
\begin{equation}
  P^s(t)=\exp [-(t-t_0)/\tau^s(Re)], 
  \label{eq:expDist}
\end{equation}
with $\tau^s$ estimated by the sample mean 
\begin{equation}
  \tau^s=\frac{1}{N^s}{\sum_{i=1}^{N^s}t_i^s}-t_0.
\end{equation}
The sample mean is effectively the maximum likelihood estimator of the scale parameter $\tau^s$~\cite{lawless2003}. $t_0$ has been determined the same way described in~\cite{Avila_JFM2010}. From Fig.~\ref{fig:probdist}(a), we find that the mean splitting time (slope of the distributions) decreases quickly with $Re$. The exact dependence of $\tau^s$ on $Re$ is shown in Fig.~\ref{fig:tau} (dark square points). It turns out that the best fit for the simulation data is a superexponential function represented by the dark dashed line. Hence $\tau^s(Re)$ only approaches infinity asymptotically as $Re$ decreases and consequently a non-zero splitting probability remains even for smaller $Re$.

The same method is applied to obtain the probability distributions of the decay events (or lifetimes), the result of which is plotted in Fig.~\ref{fig:probdist}(b). The lifetimes of localized turbulent stripes are also exponentially distributed, with the mean lifetime $\tau^d$ scaling superexponentially with $Re$ (red circles in Fig.~\ref{fig:tau}). Overall splitting and decay statistics show the same qualitative behaviour as pipe flow (see Ref.~\cite{Avila_Science2011}), suggesting that the key physical processes are largely independent of the geometry and possibly apply to canonical shear flows.

The intersection of the two curves $\tau^{s,d}(Re)$ in Fig.~\ref{fig:tau} fixes a distinct Reynolds number $Re\simeq 325$ where the mean timescales of both processes are in balance, namely, $\tau^s(Re)=\tau^d(Re)$. 
This value gives a lower bound for the critical $Re_c$ below which turbulence decays after sufficiently long time and it is also very close to earlier estimates of the critical point~\cite{Bottin_EPJB1998,Duguet_JFM2010} although those were carried out in domains with a much larger spanwise length. While characterisation of the spreading and decay processes for flows in a streamwise and spanwise long domains would require excessive computation time, the same line of argument would be applicable here. Again turbulence should become sustained once growth processes (streamwise and spanwise) outweigh the decay.

In analogy to contact processes like DP the turbulent and laminar domains can be viewed respectively as the active and passive state. In order of turbulence to survive in the system the splitting rate has to be larger than the decay rate, otherwise turbulence dies out. The onset of sustained turbulence may, in analogy to DP like systems, become sustained at a non-equilibrium phase transition~\cite{Chate_PRL1987,*Manneville_PRE2009}. As a clear signature the spacing between active sites should become scale invariant close to the critical point and hence the passive regions have no characteristic length. 


\begin{figure*}[!ht]
  \begin{center}
    \includegraphics[width=\textwidth]{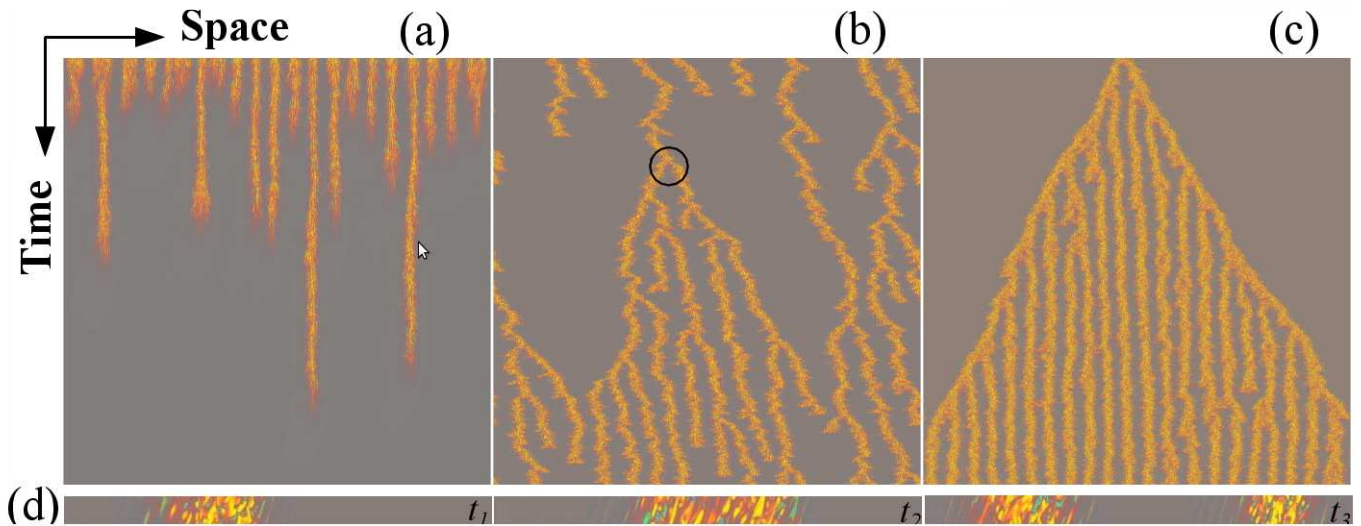}
  \end{center}
  \caption{Spatiotemporal diagrams at different regimes: (a) subcritical at $Re=300$; (b) slightly above the critical point at $Re=329.5$; (c) supercritical at $Re=360$. (d) A close-up view of the splitting event at the moment indicated by the circle in (b). Snapshots are taken at the mid $\theta$-$z$ plane. Colormap shows the streamwise vorticity.}
  \label{fig:spacetime}
\end{figure*}

To test this another set of simulations are performed and we extend the box size
in the stripe-normal direction $\hat{e}_z$ to $960h$ such that spatial correlations are taken into account. Through the simulations at three different $Re$ (subcritical, critical and supercritical), the observed dynamics agree qualitatively with the DP model (see Fig.~\ref{fig:spacetime}): At $Re=300$, the flow returns to the laminar state after sufficiently long time; At $Re=360$, an initial single stripe quickly splits until it reaches a statistically stationary state with a typical stripe spacing. At $Re=329.5$ on the other hand turbulence persists but the stripe spacings change throughout, exploring all scales permitted in the given domain size. An example of stripe splitting is shown in Fig. 4(d). After an initial increase in width, the stripe breaks up into two segments of similar size that then continue to separate. It should be noted that this process differs from puff splitting in pipes where the new puff originates from a thin filament of vorticity that disconnects from the original puff (Fig. 2A in~\cite{Avila_Science2011}). Furthermore in pipes splitting exclusively occurs in the downstream direction whereas in the present case no preferable splitting direction exists. 

\begin{figure}[!ht]
  \begin{center}
    \includegraphics[width=.45\textwidth]{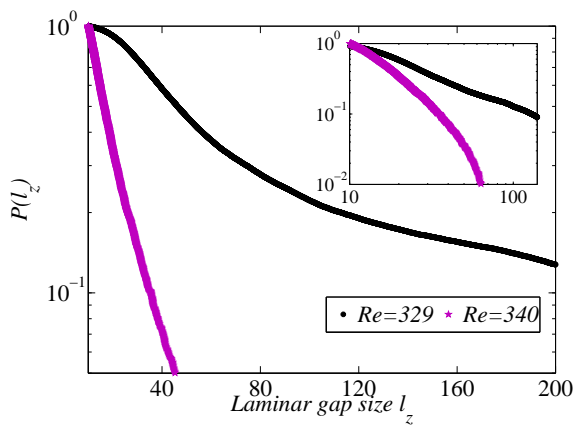}
  \end{center}
  \caption{Size distributions of laminar gaps at $Re=329$ and $Re=340$ in semi-logarithmic scale. The length in $z$ direction is $L_z=960h$. Inset is the plot in log-log scale.}
  \label{fig:lamGaps}
\end{figure}

For a quantitative evaluation of the laminar spacing, sizes of laminar gaps are measured by setting a cut-off value to the averaged vorticity, below which the flow is considered to be laminar, and the distributions are tested to be insensitive to the cut-off value (except for shifts of the absolute values). Data are sampled at the (quasi-)stationary state, over a time interval of $\mathcal{O}(10^5) \tau_h$. It is observed that the size distributions sufficiently above critical are exponential and hence possess a characteristic size. Close to the critical point the distributions follow a power law instead ($Re=329$ in Fig.~\ref{fig:lamGaps}), implying that there is no characteristic length. Size distributions hence indeed exhibit scale invariance, confirming that the intersection point in Fig.~\ref{fig:tau} marks a phase transition and the onset of sustained turbulence. Furthermore the circumstance that size distributions follow a power law close to criticality and become exponential at larger supercritical values show that the resulting intermittent flows are intrinsically irregular and do not form fixed patterns as had been proposed previously~\cite{Prigent_PRL2002,*Manneville_epl2012}.

In summary our investigation shows that the decay of turbulent stripes in Couette flow is a memoryless process and that individual stripes remain transient. Only through spatial proliferation turbulence can eventually become sustained. The splitting process that underlies this expansion of turbulence is memoryless. Overall this behaviour closely resembles the onset of turbulence in pipe flow and it is hence likely that these processes are generally responsible for the onset of turbulence in canonical shear flows. In contrast to pipe flow here the relevant mean times of the decay and spreading processes are several orders of magnitude smaller which allows measurements close to criticality. The scale invariant distributions observed show that the transition to turbulence is a non-equilibrium second order phase transition. While many aspects of this transition are analogues to directed percolation measurements of critical exponents and the universality class of the transition to turbulence would require extremely long integration times and set challenges for future studies.

We thank the Max Planck Society for supporting this work. The research leading to these results has received funding from the European Research Council under the European Union's Seventh Framework Programme (FP/2007-2013) / ERC Grant Agreement 306589. L. S. and B. H. acknowledge research funding by Deutsche Forschungsgemeinschaft (DFG) under grant SFB 963/1 (project A8). Numerical simulations were performed thanks to the CPU time allocations of GWDG in G\"ottingen and of JUROPA in J\"ulich Supercomputing Center (project HGU17). 

\bibliography{./ref_scale_invariance.bib}

\end{document}